\documentclass[copyright,creativecommons]{eptcs}

\providecommand{\event}{EXPRESS/SOS 2014}
\usepackage{breakurl}
\usepackage{enumerate}
\usepackage{hyperref}
\usepackage[active]{srcltx}
\usepackage[utf8]{inputenc}
\usepackage{mathpartir}
\usepackage{latexsym}
\usepackage{amsmath}
\usepackage{amsthm}
\usepackage{amssymb}
\usepackage{amscd}
\usepackage{macros}
\usepackage{color}
\usepackage{graphics}
\usepackage{version}
\usepackage{url}
\usepackage{soul}
\usepackage{placeins}
\usepackage{stmaryrd}
\usepackage[normalem]{ulem}
\usepackage{pdfpages}
\usepackage[nottoc]{tocbibind}
\usepackage{booktabs}

\usepackage{psi-macros}
\usepackage{todo}

\newtheorem{theorem}{Theorem}  
\newtheorem{definition}{Definition}  
\newtheorem{lemma}[theorem]{Lemma}  

\hypersetup{
    colorlinks,%
    citecolor=black,%
    filecolor=black,%
    linkcolor=black,%
    urlcolor=black
}

\title{Priorities Without Priorities: Representing Preemption in Psi-Calculi}
\author{Johannes {\AA}man Pohjola
\institute{Uppsala University, Sweden}
\email{johannes.aman-pohjola@it.uu.se}
\and
Joachim Parrow
\institute{Uppsala University, Sweden}
\email{joachim.parrow@it.uu.se}
}
\def\titlerunning{Priorities Without Priorities}
\def\authorrunning{J. {\AA}man Pohjola and J. Parrow}

\begin{document}

\maketitle

\begin{abstract}
Psi-calculi is a parametric framework for extensions of the pi-calculus with data terms and arbitrary logics. In this framework there is no direct way to represent action priorities, where an action can execute only if all other enabled actions have lower priority.
We here demonstrate that the 
psi-calculi parameters can be chosen such that the effect of action priorities can be encoded.

To accomplish this we define an extension of psi-calculi with action priorities, and show that for every   calculus in the extended framework there is a corresponding ordinary psi-calculus, without priorities, and a translation between them that satisfies strong operational correspondence. This is a significantly stronger result than for most encodings between process calculi in the literature.

We also formally prove in Nominal Isabelle that the standard congruence and structural laws about strong bisimulation hold in psi-calculi extended with priorities.



\end{abstract}

\section{Introduction}

Priorities in process calculi allow certain actions to take precedence over others. This is useful when modelling systems because it admits more fine-grained control over the model's behaviour. Phenomena that exhibit prioritised behaviour include eg.~interrupts in operating systems, and exception handling in programming languages. In this paper we demonstrate how priorities can be represented in the psi-calculi framework, by encoding them into the logical theory that determines how actions are generated by process syntax.

Psi-calculi~\cite{bengtson.johansson.ea:psi-calculi-long} is a family of applied process calculi that generalises the {\pic} in three ways. First, the subjects (designating the communication channels) and objects (designating the communicated data) of input and output actions may be \emph{terms} taken from an arbitrary set, and not just single names. Second, equality tests on names are replaced by tests of predicates called \emph{conditions}, taken from an arbitrary logic. Finally, the process syntax is extended with \emph{assertions}, which can be seen as introducing new facts about the environment in which a process executes. The unguarded assertions of a process influence the evaluation of conditions and the connectivity between channel terms, and can change as the process executes. 

In this paper, we  show that the psi-framework is sufficiently expressive to represent action priorities derived from a priority order on the communication channels. We are interested in priorities for two reasons. First, previous work on priorities indicate that they are highly expressive: Jeffrey defines a process calculus with time and priority where timed processes can be encoded in the untimed fragment of the calculus~\cite{DBLP:conf/ftrtft/Jeffrey92}; Jensen shows that CCS augmented with priority choice can encode broadcast communication~\cite{Torp94interpretingbroadcast}; and Versari et~al.~shows that CCS with priority and only the prefix and parallel operators can solve both leader election (unlike the {\pic}), and the ``last man standing''-problem (unlike the broadcast {\pic})~\cite{DBLP:conf/concur/VersariBG07}. Second, we are not aware of another process calculus (without priorities) where adding priorities has been shown to yield no increased expressiveness. The prevailing methods to introduce priorities in process algebras are through semantic rules with negative premises or new auxiliary relations to express the absence of higher-priority actions; we shall need none of those.

We accomplish our result as follows. First we define an extension of the psi-calculi framework with explicit channel priorities, where the priority level of a channel can change dynamically during process execution, as defined by an auxiliary relation representing absence of actions. We formally prove, using the interactive theorem prover Isabelle~\cite{nipkow:isabelle}, that in this setting strong bisimilarity satisfies the usual algebraic laws and congruence properties familiar from the {\pic}. We proceed to show that for every psi-calculus with priorities, separate choice and prefix-guarded replication, it can be encoded in a standard psi-calculus without priorities. This encoding satisfies particularly strong quality criteria, namely strong operational correspondence, meaning that the translation does not introduce any protocol in the target language. The main idea is that we use a non-monotonic logic for the assertions, where the appearance of enabled high-priority channels can temporarily prevent lower priority channels from resulting in actions.

The rest of the paper is structured as follows. In Section~2 we briefly recapitulate the essentials of psi-calculi, and in Section~3 we define the extension with explicit channel priorities. Section~4 contains an encoding into standard psi-calculi. In Section~5 we establish strong operational correspondence and briefly discuss other criteria for encodings, among them full abstraction, and Section~6 contains conclusions with future work.

Full proofs of all theorems presented in this paper are available online at \url{http://www.it.uu.se/research/group/mobility/prio-proofs.pdf}.




\section{Psi-calculi}\label{sec:psi}

The following is a quick recapitulation of the psi-calculi framework. For an in-depth introduction with motivations and examples we refer the reader to \cite{bengtson.johansson.ea:psi-calculi-long}.

We assume a countably infinite set of atomic \emph{names} $\,\nameset$ ranged over by $a,b,\ldots,z$. Intuitively, names are the symbols that can be scoped and be subject to substitution. 
A \emph{nominal set}~\cite{PittsAM:nomlfo-jv,Gabbay01anew} is a set equipped with a formal notion of what it means to swap names in an element; this leads to a notion of when a name $a$ occurs in an element $X$, written $a \in \names{X}$ (pronounced ``$a$ is in the support of $X$'').
We write $a \freshin X$, pronounced ``$a$ is fresh for $X$'', for $a \not\in \names{X}$, and if $A$ is a set of names we write $A\freshin X$ to mean $\forall a \in A \;.\; a \freshin X$.
In the following $\tilde{a}$ is a finite sequence of names. The empty sequence is written $\epsilon$ and the concatenation of $\tilde{a}$ and $\tilde{b}$ is written $\tilde{a} \tilde{b}$.
We say that a function symbol is \emph{equivariant} if all name swappings distribute over it.
 

A \emph{nominal datatype} is a nominal set together with a set of 
functions on it. In particular we shall consider substitution functions that
substitute elements  for  names. If $X$ is an element of a datatype,
$\tilde{a}$ is a sequence of names without duplicates and $\tilde{Y}$ is an
equally long sequence of elements of possibly another datatype, 
the \emph{substitution}
$X\subst{\tilde{a}}{\tilde{Y}}$ is an element of the same datatype as
$X$. 
The substitution function can be chosen freely, but must satisfy certain natural laws regarding the treatment of names; it must be equivariant, the names $\vec{a}$ in $X[\vec{a}:=\vec{T}]$ must be alpha-convertible as if they were binding in $X$. See~\cite{bengtson.johansson.ea:psi-calculi-long} for details.

A psi-calculus is defined by instantiating three nominal data types  and four equivariant operators; formally it is a tuple  $(\terms,\assertions,\conditions,\vdash,\otimes,\sch,\unit)$ as follows.
\begin{definition}[Psi-calculus parameters]
\label{def:parameters}
A psi-calculus requires the three (not necessarily disjoint) nominal data types:
\iffalse 
\[\begin{array}{ll}
\terms & \mbox{the (data) terms, ranged over by $M,N$} \\
\conditions  & \mbox{the conditions, ranged over by $\varphi$}\\
\assertions & \mbox{the assertions, ranged over by $\Psi$}
\end{array}\]
\else
the (data) terms $\terms$, ranged over by $M,N$,
the conditions $\conditions$, ranged over by $\varphi$, 
the assertions $\assertions$, ranged over by $\Psi$,
\fi 
and the four operators:
\[\begin{array}{llcll}
{\sch}\in  \terms \times \terms \to \conditions & \mbox{Channel Equivalence} & \quad &
{\ftimes}\in \assertions \times \assertions \to \assertions& \mbox{Composition} \\
\one: \assertions& \mbox{Unit} & \quad &
{\vdash}\subseteq \assertions \times \conditions & \mbox{Entailment} \\
\end{array}
\]
\end{definition}

The binary functions above will be written in infix. 
Thus, $M \sch N$ is a condition, pronounced ``$M$ and $N$ are channel equivalent''.
We write $\Psi \vdash \varphi$, pronounced ``$\Psi$ entails $\varphi$'', for $(\Psi, \varphi)\in\; \vdash$,
and if $\Psi$ and $\Psi'$ are assertions then so is $\Psi \ftimes \Psi'$, which intuitively represents the conjunction of the information in $\Psi$ and $\Psi'$.

We say that two assertions are \emph{statically equivalent}, written $\Psi \sequivalent \Psi'$ if they entail the same conditions, i.e. for all $\varphi$ we have that $\Psi \vdash \varphi$ iff $\Psi' \vdash \varphi$.  
We impose certain requisites on the sets and operators: channel equivalence must be symmetric and transitive, $\ftimes$ must be
compositional with regard to $\sequivalent$, and the assertions with
$(\ftimes,\one)$ form an abelian monoid modulo $\sequivalent$. Finally, substitution $M\subst{\ve{a}}{\ve{T}}$ on terms must be such that if the names $\ve{a}$ are in the support of $M$, the support of $\ve{T}$ must be in the support of $M\subst{\ve{a}}{\ve{T}}$.

A \emph{frame} is an assertion together with a sequence of names that bind into it: it is of the form $\framepair{\frnames{}}{\Psi}$ where $\frnames{}$ binds into the assertion~$\Psi$. We use $F,G$ to range over frames.
We overload $\Psi$ to also mean~$\framepair{\epsilon}{\Psi}$
and $\ftimes$ to composition on frames defined by $\framepair{\frnames{1}}{\Psi_1} \ftimes \framepair{\frnames{2}}{\Psi_2} = 
\framepair{\frnames{1} \frnames{2}}{(\Psi_1 \ftimes \Psi_2)}$ where 
$\frnames{1}\freshin\frnames{2},\Psi_2$
 and vice versa. 
We write
$\Psi \ftimes F$ to mean $\framepair{\epsilon}{\Psi} \ftimes F$, and
 $(\nu
c)(\framepair{\frnames{}}{\Psi})$ for~$\framepair{c\frnames{}}{\Psi}$.

We define $F \vdash \varphi$ to mean that there exists an alpha variant  $\framepair{\frnames{}}{\Psi}$ of $F$ such that  $\frnames{} \freshin \varphi$ and $\Psi \vdash \varphi$. We also define~\mbox{$F\sequivalent G$} to mean that for all $\varphi$ it holds that $ F \vdash \varphi$ iff $ G \vdash \varphi$.

\begin{definition}[Psi-calculus agents]\label{def:agents}
Given a psi-calculus $\mathcal{P}$ with parameters as in Definition~\ref{def:parameters}, 
the \emph{agents} $\processes(\mathcal{P})$, ranged over by $P,Q,\ldots$,  are of the following forms.

{\rm
\[
\begin{array}{ll}
\nil                         & \mbox{Nil} \\
\outlabel{M}{N} \sdot P     & \mbox{Output}\\
\inprefix{M}{\ve{x}}{N} \sdot P     & \mbox{Input}\\
\case{\ci{\varphi_1}{P_1}\casesep\cdots\casesep\ci{\varphi_n}{P_n}}\quad
&\mbox{Case} \\
(\nu a)P                      & \mbox{Restriction}\\
P \pll Q                      & \mbox{Parallel}\\
! P                           & \mbox{Replication} \\
\pass{\Psi}                        & \mbox{Assertion}
\end{array}\]
}

\noindent Restriction $(\nu a)P$ binds $a$ in $P$ and input $\inprefix{M}{\ve{x}}{N} \sdot P$ binds $\vec{x}$ in both $N$ and $P$.
An occurrence of a subterm in an agent is \emph{guarded} if it is a proper subterm of an input or output term.
An agent is \emph{assertion guarded} if it contains no unguarded assertions. 
An agent is \emph{well-formed} if in $\inprefix{M}{\ve{x}}N.P$  it holds that $\ve{x} \subseteq \names{N}$ is a sequence without duplicates, 
that  in a replication $!P$ the agent $P$ is assertion guarded, 
and that in $\case{\ci{\varphi_1}{P_1}\casesep\cdots\casesep\ci{\varphi_n}{P_n}}$ the agents  $P_i$ are assertion guarded.
\end{definition}
The agent
$\case{\ci{\varphi_1}{P_1}\casesep\cdots\casesep\ci{\varphi_n}{P_n}}$ 
is sometimes abbreviated as
\mbox{\rm $\case{\ci{\ve{\varphi}}{\ve{P}}}$}.
 We sometimes write $\inprefixempty{M}(x).P$ for $\inprefix{M}{x}{x}.P$.
From this point on, we only consider well-formed agents.

 The \emph{frame $\frameof{P}$ of an agent} P is defined inductively as follows:
\begin{mathpar}
\frameof{\inprefix{M}{\ve{x}}{N} \sdot P} = 
\frameof{\outprefix{M}{N} \sdot P} =
\frameof{\nil} = 
\frameof{\case{\ci{\ve{\varphi}}{\ve{P}}}} = 
\frameof{!P} = \one \and
\frameof{\pass{\Psi}} = \framepair{\epsilon}{\Psi} \and
\frameof{P \pll Q} = \frameof{P} \ftimes \frameof{Q} \and
\frameof{\res{b}P} = (\nu b)\frameof{P}
\end{mathpar}

The \emph{actions} ranged over by $\alpha, \beta$ are of the following three kinds:\\
\emph{Output} $\outlabel{M}{(\nu \tilde{a})N}$, 
\emph{input} $\inlabel{M}{N}$, and \emph{silent} $\tau$. Here we refer to $M$ as the \emph{subject} and $N$ as the \emph{object}. We define 
$\bn{\outlabel{M}{(\nu \tilde{a})N}} = \tilde{a}$, and $\bn{\alpha}=\emptyset$
if $\alpha$ is an input or $\tau$.
As in the pi-calculus, the output $\outlabel{M}{(\nu \tilde{a})N}$ represents
an action sending $N$ along $M$ and opening the scopes of the names
$\tilde{a}$. 

\begin{table*}[tb]

\begin{mathpar}

\inferrule*[Left=\textsc{In}]
    {\Psi \vdash K \sch M}   
{\framedtrans{\Psi}{\inprefix{M}{\ve{y}}{N} \sdot
P}{\inlabel{K}{N}\subst{\ve{y}}{\ve{L}}}{P\subst{\ve{y}}{\ve{L}}}}


\inferrule*[left=\textsc{Out}]
    {\Psi \vdash M \sch K}
    {\framedtrans{\Psi}{\outprefix{M}{N} \sdot P}{\outlabel{K}{N}}{P}}


\inferrule*[left={\textsc{Case}}]
    {\framedtrans{\Psi}{P_i}{\alpha}{P'} \\ \Psi \vdash \varphi_i}
    {\framedtrans{\Psi}{\case{\ci{\ve{\varphi}}{\ve{P}}}}{\alpha}{P'}}

\inferrule*[left=\textsc{Par},  right={$
    \bn{\alpha} \freshin Q
$}] 
{\framedtrans{\frass{Q} \ftimes \Psi}{P} {\alpha}{P'}}
{\framedtrans{\Psi}{P \pll Q}{\alpha}{P' \pll Q}}

\inferrule*[left=\textsc{Com}, right={$
    \ve{a} \freshin Q    
$}]
 {\Psi \ftimes \frass{P} \ftimes \frass{Q} \vdash M \sch K \\
 \framedtrans{\frass{Q} \ftimes \Psi}{P}{\outlabel{M}{(\nu \ve{a})N}}{P'} \\
  \framedtrans{\frass{P} \ftimes \Psi}{Q}{\inlabel{K}{N}}{Q'} 
  }
       {\framedtrans{\Psi}{P \pll Q}{\tau}{(\nu \ve{a})(P' \pll Q')}}

\inferrule*[left=\textsc{Rep}]
   {\framedtrans{\Psi}{P \pll !P}{\alpha}{P'}}
   {\framedtrans{\Psi}{!P}{\alpha}{P'}}

\inferrule*[left=\textsc{Scope}, right={$b \freshin \alpha,\Psi$}]
    {\framedtrans{\Psi}{P}{\alpha}{P'}}
    {\framedtrans{\Psi}{(\nu b)P}{\alpha}{(\nu b)P'}}

\inferrule*[left=\textsc{Open}, right={$\inferrule{}{b \freshin \ve{a},\Psi,M\\\\b \in \names{N}}$}]
    {\framedtrans{\Psi}{P}{\outlabel{M}{(\nu \ve{a})N}}{P'}}
    {\framedtrans{\Psi}{(\nu b)P}{\outlabel{M}{(\nu \ve{a} \cup \{b\})N}}{P'}}

\end{mathpar}
\caption{\rm Structured operational semantics. Symmetric versions of \textsc{Com} and \textsc{Par} are elided. In the rule $\textsc{Com}$ we assume that
$\frameof{P} =
\framepair{\frnames{P}}{\Psi_P}$ and   $\frameof{Q} =
\framepair{\frnames{Q}}{\Psi_Q}$ where $\frnames{P}$ is fresh for all of 
$\Psi, \frnames{Q}, Q, M$ and $P$, and that $\frnames{Q}$ is similarly fresh. In the rule
\textsc{Par} we assume that $\frameof{Q} = \framepair{\frnames{Q}}{\Psi_Q}$
where $\frnames{Q}$ is fresh for
$\Psi, P$ and $\alpha$. 
In $\textsc{Open}$ the expression $\tilde{a} \cup \{b\}$ means the sequence
$\tilde{a}$ with $b$ inserted anywhere.
}
\label{table:full-struct-free-labeled-operational-semantics}
\end{table*}

\FloatBarrier
\begin{definition}[Transitions]
\label{transitions}
A \emph{transition} is written \mbox{$\framedtrans{\Psi}{P}{\alpha}{P'}$},
meaning that in the environment $\Psi$, $P$ can do $\alpha$ to become $P'$.  The transitions are defined inductively in 
Table~\ref{table:full-struct-free-labeled-operational-semantics}.
We abbreviate $\unit \frames \trans{P}{\alpha}{P'}$ as $\trans{P}{\alpha}{P'}$.
\end{definition}


We identify alpha-equivalent agents, frames and transitions. In a transition the names in $\bn{\alpha}$ bind into both the action object and the derivative, therefore $\bn{\alpha}$ is in the support of $\alpha$ but not in the support of the transition.

\FloatBarrier

\begin{definition}[Strong bisimulation]\label{defn:strongbisim}
A \emph{strong bisimulation}
 $\cal R$ is a ternary relation on assertions and pairs of agents such that
 ${\cal R}(\Psi,P,Q)$ implies
 \begin{enumerate}\addtolength{\itemsep}{0.3\baselineskip}
 \item Static equivalence:
  $\Psi \ftimes \frameof{P} \sequivalent \Psi \ftimes \frameof{Q}$; and
 \item Symmetry: ${\cal R}(\Psi,Q,P)$; and
 \item Extension of arbitrary assertion:
   $\forall \Psi'.\ {\cal R}(\Psi \ftimes \Psi',P,Q)$; and
 \item Simulation:
   for all $\alpha, P'$ such that $\framedtrans{\Psi}{P}{\alpha}{P'}$ and
   $\bn{\alpha}\freshin \Psi,Q$,\\there exists $Q'$ such
   that $\framedtrans{\Psi}{Q}{\alpha}{Q'}$ and ${\mathcal R}(\Psi , P', Q')$.
\end{enumerate}
 \label{def:bisim}
We define $\Psi \frames P \bisim Q$ to mean that there exists a bisimulation ${\cal R}$ such that
${\cal R}(\Psi,P,Q)$, and write $P \bisim Q$, pronounced $P$ and $Q$ are {\em (strongly) bisimilar}, for $\unit \rhd P \bisim Q$.
\end{definition}

\begin{definition}[Strong congruence]\label{def:strongcong}
We define $P \sim_\Psi Q$ to mean that for all substitution sequences $\sigma$, $\Psi \frames P\sigma \bisim Q\sigma$ holds. We write $P \sim Q$, pronounced $P$ is {\em (strongly) congruent} to $Q$,  to mean $P \sim_\unit Q$.
\end{definition}

We have shown~\cite{bengtson.johansson.ea:psi-calculi-long} that strong bisimilarity preserves all operators except input, and that strong congruence is a congruence and satisfies the expected algebraic laws for structural congruence.

\section{Extension: Psi-calculi with priorities}

The most common approaches to implementing priorities in process calculi are (1) to add a priority operator $\Theta$ such that $\Theta(P)$ may only take the highest-priority actions of $P$ as defined by some ordering on actions~\cite{BBK86}, and (2) to alway enforce priorities, rather than only at special operators~\cite{DBLP:conf/lics/CleavelandH88,HPA2001}. In order to avoid introducing a new operator, we follow the second approach.

We associate a priority level to actions that may depend on the assertion environment, and hence change dynamically as a process evolves.
The intuition is that we write $\Psi \vdash \Prio{M}{p}$ to mean that the priority level of communication on the channel $M$ in the environment $\Psi$ is $p$, where lower values of $p$ indicate \emph{higher} priority.
Priorities are subject to some natural constraints: they must be equivariant, and in a given assertion, channel equivalent terms must have the same unique priority level.

\begin{definition}[Psi-calculi with priorities]\label{defn:psiwithprio} A \emph{psi-calculus with priorities}, ranged over by $\mathcal{P},\mathcal{Q}$, is a tuple $(\terms,\assertions,\conditions,\vdash,\otimes,\sch,\unit,\PRIO)$ such that
\begin{enumerate}
  \item $(\terms,\assertions,\conditions,\vdash,\otimes,\sch,\unit)$ is a psi-calculus, and
  \item${\PRIO}\;$  of type $\terms\times\numbers \Rightarrow \conditions$ is an equivariant operator written in infix, i.e., we write $\Prio M p$ for $\PRIO(M,p)$,
  such that for all $\Psi,M,N$, if $\Psi \vdash M \sch N$ then there is a unique $p \in \numbers$ such that $\Psi \vdash \Prio M p$ and $\Psi \vdash \Prio N p$.
\end{enumerate}
\end{definition}


The semantics of psi-calculi with priorities is as the semantics of psi-calculi, but with two changes. The first is that $\tau$ actions are replaced with $\tau:p$ actions, where $p$ is the priority level of the transition. The second is that the rules are augmented with side conditions that prevent a process from taking low priority actions. This has a natural formulation in terms of negative premises~\cite{DBLP:journals/jacm/BolG96}, but in order to make implementation in Isabelle easier we instead define the semantics in two layers, following~\cite{DBLP:conf/lics/CleavelandH88,HPA2001,DBLP:conf/esop/Versari07}.

The bottom layer is denoted with the transition arrow $\prionegtransarrow{}$ and is used to determine which transitions would be available, disregarding priorities. The semantics of $\prionegtransarrow{}$ is exactly as in Table~\ref{table:full-struct-free-labeled-operational-semantics} with the sole extension that the \textsc{Com} rule generates an action of kind $\tau:p$, where $p$ is derived from the priority of the channel. 
We then define a predicate $\priook(\alpha,\Psi,P)$, which intuitively means that no $\tau$ transition whose priority is higher than that of $\alpha$ can be derived from $P$ in $\Psi$. Finally we define $\priotransarrow{}$ to represent 
transitions respecting priorities, where the \textsc{Case}, \textsc{Par}, and \textsc{Com} rules get side conditions using $\priook$.

\begin{definition}
\[\priook(\alpha,\Psi,P) \defn
  \neg \exists n\;P'. (\Psi\frames {P\prionegtransarrow{\tau:n}}\; P'\land n <\Prioof{(\Psi\otimes\frameof{P},\alpha)})
\]

where $\Prioof{(F,\alpha)}$ is defined to be $p$ if either $\alpha = \tau:p$ or $F \vdash \Prio {\subj{\alpha}} p$, and $\prionegtransarrow{}$ is defined in
Definition~\ref{priotransitions} below.
\end{definition}

\begin{definition}[Transitions with priorities]\label{priotransitions} The transitions of psi-calculi with priorities are defined inductively by the same rules as in Table~\ref{table:full-struct-free-labeled-operational-semantics}, but with all occurrences of $\apitransarrow{}$ replaced with $\priotransarrow{}$, and the $\textsc{Case}$, $\textsc{Com}$ and $\textsc{Par}$ rules replaced by the following:

\begin{mathpar}

\inferrule*[left={\textsc{Case}},right={$\priook (\alpha,\Psi,\case{\ci{\ve{\varphi}}{\ve{P}}})$}]
    {\prioframedtrans{\Psi}{P_i}{\alpha}{P'} \\ \Psi \vdash \varphi_i}
    {\prioframedtrans{\Psi}{\case{\ci{\ve{\varphi}}{\ve{P}}}}{\alpha}{P'}}

\inferrule*[left=\textsc{Par},  right={$
  {\begin{array}{l}
    \priook (\alpha,\Psi,P\mid Q)\\
    \bn{\alpha} \freshin Q
  \end{array}
}$}] 
{\prioframedtrans{\frass{Q} \ftimes \Psi}{P} {\alpha}{P'}}
{\prioframedtrans{\Psi}{P \pll Q}{\alpha}{P' \pll Q}}

\inferrule*[left=\textsc{Com}, right={$
  {\begin{array}{l}
    \priook (\tau:p,\Psi,P\mid Q)\\
    \ve{a} \freshin Q    
  \end{array}
}$}]
 {\Psi \ftimes \frass{P} \ftimes \frass{Q} \vdash M \sch K\qquad\Psi \otimes \frass{P} \otimes \frass{Q}\vdash\Prio M p \\
 \prioframedtrans{\frass{Q} \ftimes \Psi}{P}{\outlabel{M}{(\nu \ve{a})N}}{P'} \\
  \prioframedtrans{\frass{P} \ftimes \Psi}{Q}{\inlabel{K}{N}}{Q'} 
  }
       {\prioframedtrans{\Psi}{P \pll Q}{\tau:p}{(\nu \ve{a})(P' \pll Q')}}
\end{mathpar}

The transition relation $\prionegtransarrow{}$ is defined by the same rules as $\priotransarrow{}$, but with all side conditions involving $\priook$ omitted.
\end{definition}

Strong bisimulation and strong congruence on psi-calculi with priorities can be obtained from Definitions~\ref{defn:strongbisim}-\ref{def:strongcong} by replacing all occurrences of $\apitransarrow{}$ with $\priotransarrow{}$. The meta-theory pertaining to strong bisimulation from the original psi-calculi carries over to psi-calculi with priorities, and formal proofs in Isabelle have been carried out:

\begin{theorem}\label{thm:priobisim} Strong congruence $\sim$ on psi-calculi with priorities is a congruence, and satisfies
\[\begin{array}{rcll}
{P} &\sim& {P \parop \nil} & \\
{P\parop ( Q \parop R)}&\sim&{(P \parop Q) \parop R} & \\
{P \parop Q}&\sim&{Q \parop P} & \\
{(\nu a)\nil}&\sim&{\nil} & \\
{(\nu a)(\nu b)P}&\sim&{(\nu b)(\nu a)P} & \\
{!P}&\sim&{P \parop !P} & \\
{P \parop (\nu a) Q} &\sim& {(\nu a)(P \parop Q)} & \mbox{if $a \freshin P$}\\
{\outprefix{M}{N}.(\nu a)P} &\sim& {(\nu a)\outprefix{M}{N}.P} & \mbox{if $a \freshin M,N$}\\
{\inprefix{M}{\vec{x}}{N}.(\nu a)P} &\sim& {(\nu a)\inprefix{M}{\vec{x}}{N}.P} & \mbox{if $a \freshin M,\ve{x},N$} \\
{\caseonly\;{\ci{\vec{\varphi}}{\vec{(\nu a)P}}}} &\sim& {(\nu a)\caseonly\;{\ci{\vec{\varphi}}{\vec{P}}}} & \mbox{if $a \freshin \ve{\varphi}$}
\end{array}\]

\end{theorem}

As an example, Versari's $\pi @$~\cite{DBLP:conf/esop/Versari07} is an extension of the {\pic} with priorities. Input and output prefixes in $\pi @$ are of form $\mu : k(y)$ and $\overline{\mu} : k\langle z\rangle$, where $\mu$ is the subject, $k$ is the priority level and $y$ and $z$ are the objects. The semantics is the standard reduction semantics of the {\pic}, augmented with side conditions stating that no higher-priority reduction is possible, similar to our use of the $\priook$ predicate.

$\pi @$ can be recovered in our framework as follows. For simplicity we consider only monadic synchronisation. Let the terms be the union of $\nameset$ (corresponding to objects in $\pi @$) and $\{a:n | a \in \nameset, n \in \numbers\}$ (corresponding to subjects annotated with their priority level), let the conditions be the booleans and the assertions be $\{\unit\}$. Define channel equivalence and $\PRIO$ so that $a:n$ is equivalent to itself and has priority $n$.



As an immediate consequence, we equip $\pi @$ with a labelled semantics and a theory of strong bisimulation; no labelled semantics or bisimulation theory has been previously developed for $\pi @$.

Note that in our representation of $\pi @$, it is possible to write agents where the term $a:n$ occurs in object position. We can rule out such ill-formed agents by using the sort system for psi-calculi described in~\cite{Borgstr_m_2014}, the details of which are beyond the scope of the present paper.

For a slightly more involved example, we consider dynamic priorities. We define a psi-calculus with priorities based on the \pic, with the addition that channels may have one of two priority levels: $0$ (high) and $1$ (low). Rather than annotating prefixes with a priority level, we let channels have high priority by default, and let our assertions be the set of channels whose priority have been flipped to low priority. If a channel is asserted to be flipped twice, the assertions cancel each other and the channel is flipped back to high priority. Thus we may flip the priority of a channel $a$ dynamically by asserting $\Set{a}$. Similarly, asserting $\Set{a,b}$ flips the priorities of both $a$ and $b$. Composition of assertions is exclusive or, e.g.~$\Set{a}\otimes\Set{a,b} = \Set{b}$. To illustrate how this calculus can be used, suppose we want to enforce a fairness scheme such that synchronisations on two channels $x$ and $y$ are guaranteed to interleave. This can be achieved by swapping the priorities of $x$ and $y$ after every such synchronisation, as in the following derivation sequence, where for all $z\in\Set{x,y}$ we let $P_z = \pass{\Set z} \pll !\overline{x}.\pass{\Set{x,y}} \pll !\overline{y}.\pass{\Set{x,y}}$.
\[
\begin{array}{rclcl}
\one & \frames & 
    P_y \pll x\sdot x\sdot x \pll y
    & \priotransarrow{\tau:0} & P_x \pll x\sdot x \pll y \\
    & & & \priotransarrow{\tau:0} & P_y \pll x\sdot x \\
    & & & \priotransarrow{\tau:0} & P_x \pll x \\
    & & & \priotransarrow{\tau:1} & P_y
\end{array}
\]
Note that the above $\tau$ sequence is the only possible $\tau$ sequence --- as long as both $x$ and $y$ are available they are guaranteed to be consumed alternatingly.

Formally, we define this psi-calculus by letting
$\terms = \nameset$, $\conditions = \Set{x = y \;|\; x, y \in \terms} \cup \Set{\Prio M n \;|\; M \in \terms \wedge n \in \numbers}$ and by letting $\assertions$ be the finite sets of names. Moreover, let $\one$ be the empty set and $A \ftimes B = (A \cup B) - (A \cap B)$. Entailment is defined so that $\Psi \vdash x=y$ iff $x=y$, $\Psi \vdash \Prio x 1$ iff $x \in \Psi$, and $\Psi \vdash \Prio x 0$ iff $x \not\in \Psi$. Finally, we let channel equivalence be syntactic equality on names.

The definition of composition as the pairwise exclusive or on the elements of its arguments achieves the priority flip in a manner that is associative, commutative and compositional. This is a useful general technique for constructing psi-calculi where facts can be retracted.

\section{Encoding priorities}

In this section we present a translation from psi-calculi with priorities to the original psi-calculi. The main idea is that we augment the assertions with information about prefixes, and ensure that the frame of a process records precisely its enabled prefixes. The $\priook$ predicate is thus obtained from the entailment relation. 

The main technical complication with this idea is that when $P$ takes a transition to $P'$, some of the top-level prefixes of $P$ may be absent in $P'$. The frame of $P'$ will always be the frame of $P$ composed with assertions that are guarded in $P$ and unguarded in $P'$; in other words $\frameof{P'} \simeq (\nu \frnames{P'})(\frass{P} \otimes \Psi)$. It follows that composing with this $\Psi$ must in effect retract the prefixes no longer available in $P'$ from $\frass{P}$. For this purpose we use a non-monotonic logic, where assertions contain 
 multisets with negative occurrence~\cite{blizard1990}.

\subsection{Preliminaries: integer-indexed multisets}

Intuitively, an integer-indexed multiset is like a regular multiset, except that the number of occurrences of an element may be negative.  We  use  \emph{finite integer-indexed multisets with a maximum element} (henceforth abbreviated \emph{FIMM}), ranged over by $E$. Let $\znumbers^\infty$ denote $\znumbers \cup \Set{\infty}$. Formally, the FIMMs over a set $S$ is the set of functions $E : \msetof{S}$ such that for all but finitely many elements $s \in S$, $E(s) = 0$. We define some of the usual operations on sets as follows:

\begin{mathpar}
 x \in E  \;\defn\; E(x) > 0 \and
 \emptyset \;\defn\; \lambda x. 0 \and
 E \cup E' \;\defn\; \lambda x. (E(x) + E'(x))
\end{mathpar}

The maximal element $\infty$ will be used to represent prefixes under a replication operator (these are permanently enabled and cannot ever be retracted). We will write $\{(z_0)x_0,\,\dots,\,(z_n)x_n\}$ for the multiset $E$ such that $E(x_i) = z_i$ if $0 \leq i \leq n$, and $E(x_i) = 0$ otherwise. We will sometimes write $x_i$ to mean $(1)x_i$ and $-x_i$ to mean $(-1)x_i$.


\subsection{Preliminaries: Requisites and guarding elements}\label{sec:guardelm}

From this point in the paper, we restrict attention to psi-calculi with separate choice and prefix-guarded replication. In other words, case statements have the form $\caseonly\;{\ci{\ve{\varphi}}{\ve{\alpha. P}}}$, where either every $\alpha_i$ is an input, or every $\alpha_i$ is an output. Moreover, replications are of the form $\bang \alpha.P$. These restrictions significantly simplify our definitions and proofs. In the conclusion we briefly discuss what would be involved to lift them. 

We also require that substitution has no effect on terms where the names being substituted do not occur, i.e. that if $\ve{x} \freshin M$ then $M\subst{\ve x}{\ve T} = M$. This natural requirement on substitution is found in the original publication on psi-calculi~\cite{LICS2009:Psi-calculi}, but is often omitted since it is not needed for the standard structural and congruence properties of bisimulation. 

Further, for convenience we will assume that the psi-calculus under consideration has a condition $\top$ that is always true in every context, i.e. it is such that $\forall \Psi. \Psi \vdash \top$, $\forall \sigma. \top\sigma = \top$ and $\supp{\top} = \emptyset$. If such a condition is absent, it can simply be added.

A \emph{guarding element} is simply a prefix guarded by a condition. Enriching the assertions with FIMMs of guarding elements will provide all the information necessary to encode $\priook$ in the entailment relation.

\begin{definition}[guarding elements]
The set of \emph{guarding elements} of a psi-calculus $\mathcal{P} = (\terms,\assertions,\conditions,\vdash,\otimes,\sch,\unit)$ is denoted $\prefixesof{\mathcal{P}}$ and defined as

\[
\prefixesof{\mathcal P} = \conditions \times (\{\outprefix{M}{N}: M,N \in \terms\} \cup \{\inprefix{M}{\vec{x}}{X}: M,N \in \terms\})
\]

We consider guarding elements as implicitly quotiented by alpha-equivalence, where the names $\vec{x}$ in the input prefix $\inprefix{M}{\vec{x}}{X}$ bind into $N$. We will sometimes write $\alpha$ to mean $(\top,\alpha)$.
\end{definition}

\subsection{The encoding}\label{sec:cuteencoding}

Assume a psi-calculus with priorities $\mathcal{P} = (\terms,\assertions,\conditions,\vdash,\otimes,\sch,\unit,\PRIO)$. We shall encode it in the psi-calculus $\mathcal{Q} = (\terms,\assertions',\conditions',\vdash',\otimes',\sch',(\unit,\emptyset))$, whose parameters are defined as follows:

\[
\begin{array}{rcl}
  \assertions' & = & \assertions \times (\msetof{\prefixesof{\mathcal{P}}}) \\
  \conditions' & = & \conditions \uplus (\znumbers^\infty \times \prefixesof{\mathcal{P}}) \uplus \{M \sch' N: M,N \in \terms\} \\
  (\Psi,E) \otimes' (\Psi',E') & = & (\Psi \otimes \Psi', E \cup E') \\
\end{array}
\]
\[
\begin{array}{rcl}
  (\Psi,E) \vdash' \varphi & = & \Psi \vdash \varphi \qquad \mbox{if $\varphi \in \conditions$}\\
  (\Psi,E) \vdash' (z)(\varphi, \alpha) & = & E(\varphi, \alpha) = z \\
  (\Psi,E) \vdash' M \sch' N & = & \Psi \vdash M \sch N \wedge \neg\exists M'\;N'\;n\;m\;X\;K\;\ve{x}\;\ve{L}\;\varphi\;\varphi'. \Psi \vdash M' \sch N' \\
  & & \wedge \Psi \vdash \Prio{M}{m} \wedge \Psi \vdash \Prio{M'}{n} \wedge n < m \wedge (\varphi,\inprefix{M'}{\ve x}X) \in E \\
  & & \wedge (\varphi',\outprefix{N'}{K}) \in E \wedge K = X\subst{\ve x}{\ve L} \wedge \Psi \vdash \varphi \wedge \Psi \vdash \varphi'\\
\end{array}
\]

Assertions in $\assertions'$ augment the original assertions with FIMMs of guarding elements, representing the top-level prefixes of a process. The conditions are augmented with multiplicity tests on elements of the FIMMs (only needed for technical reasons concerning the compositionality of $\otimes'$), as well as channel equivalence statements. Composition and entailment of multiplicity tests and conditions in $\conditions$ are straightforward. The definition of entailment of channel equivalence statements intuitively means that two channels $M,N$ are equivalent in $(\Psi,E)$ if (1) they are equivalent in $\Psi$, and (2) $E$ does not contain prefixes that can communicate with each other with a priority higher than that of $M,N$. This is the mechanism by which we prevent lower-priority actions in the translations: those actions that would be ruled out by $\priook$ in $\mathcal{P}$ are ruled out in $\mathcal{Q}$ by not being channel equivalent to anything.

In order to avoid bogging down the notation with brackets, we introduce some syntactic sugar for assertions in $\assertions'$. We will sometimes write $\Psi$ for $(\Psi,\emptyset)$ and $E$ for $(\unit,E)$. Further, we will sometimes write single-element multisets without the curly brackets, ie. $(z)x$ for $\{(z)x\}$. For an example, combined with the previously introduced syntactic sugar for multisets and guarding elements, we may write $(\unit,\{(1)(\top,\alpha)\})$ as simply $\alpha$, and $(\unit,\{(-1)(\top,\alpha)\})$ as $-\alpha$.

\begin{lemma} $\mathcal{Q}$ is a psi-calculus, meaning that it satisfies the requisites outlined in Section~\ref{sec:psi}.
\end{lemma}

The translation of agents from $\mathcal{P}$ to $\mathcal{Q}$ is defined by the function $\semb{\_} : \processesof{\mathcal P} \Rightarrow \processesof{\mathcal Q}$. The main idea is that in parallel to every prefix, we add the prefix as an assertion (recall that $\pass{\Psi}$ denotes the assertion $\Psi$ occurring as a process), so that it can be used when deciding channel equivalences. The continuation after the prefix contains the same prefix negatively, and since $\{\alpha\} \cup \{-\alpha\}=\emptyset$ the effect is to retract the prefix from the frame once it has been used,
 and thus ensures that the frame of an agent $\semb{P}$ contains an up-to-date copy of the top-level prefixes of $P$. Since replicated prefixes are permanently enabled, a replicated prefix is asserted with infinite multiplicity to ensure that it is never retracted. For $\caseonly$ statements, we make sure to retract the guarding elements associated with the other branches after a particular branch has been chosen.

\[
\begin{array}{rcl}
  \semb{\nil} & = & \nil \\
  \semb{\pass{\Psi}} & = & \pass{(\Psi,\emptyset)}\\
  \semb{P \parop Q} & = & \semb{P} \parop \semb{Q} \\
  \semb{(\nu x)P} & = & (\nu x)\semb{P} \\
  \semb{\alpha. P} & = & \pass{\alpha} \parop \alpha. (\semb{P} \parop \pass{-\alpha}) \\
  \semb{\bang \alpha. P} & = & \pass{(\infty)\alpha} \parop \bang \alpha. (\semb{P} \parop \pass{-\alpha}) \\
  \semb{\caseonly\;{\ci{\ve{\varphi}}{\ve{\alpha. P}}}} & = & \pass{(\ve{\varphi},\ve{\alpha})} \parop \caseonly\;{\ci{\ve{\varphi}}{\ve{\alpha}. (\ve{\semb{P}} \parop \pass{(-1)(\ve{\varphi},\ve{\alpha})})}} \\
\end{array}
\]

Recall that we require that substitution has no effect on terms where the names being substituted do not occur. To see why, consider the encoding of the input prefix $\alpha = \inprefix{M}{\ve{x}}{N}$, where $\ve{x}$ is chosen to be fresh in $M$. If the encoding takes a transition $\trans{\semb{\alpha}}{\inlabel{M}{N\subst{\ve{x}}{\ve{L}}}}{\pass{\alpha} \parop \pass{-\alpha\subst{\ve{x}}{\ve{L}}}}$, we need that $\alpha\subst{\ve{x}}{\ve{L}} = \alpha$ to achieve a retraction of $\alpha$. This follows from our requirement since $\ve{x}$ does not occur freely in $\alpha$.

\section{Quality of the encoding}\label{sec:quality}

In this section, we show that the encoding presented in Section~\ref{sec:cuteencoding} satisfies strong operational correspondence, and briefly discuss two other quality criteria: Gorla's framework~\cite{DBLP:conf/concur/Gorla08} and full abstraction.

Let $\equiv$, pronounced \emph{structural congruence}, be the smallest congruence on processes that satisfies the commutative monoid laws with respect to $(\parop,\nil)$ and the rules $!P \equiv P \parop !P$ and $\nil \equiv \pass{\one}$ and $\pass{\Psi} \parop \pass{\Psi'} \equiv \pass{\Psi \otimes \Psi'}$.

The main result of this paper is a one-to-one transition correspondence between agents in $\mathcal{P}$ and their encodings in $\mathcal{Q}$:

\begin{theorem}[Strong operational correspondence]\label{thm:stopcon}$\,$
  \begin{enumerate}
    \item If $\prioframedtrans{\Psi}{P}{\alpha}{P'}$ and $\bn\alpha \freshin P$ and $\alpha \neq \tau:p$,
      then there exists $P''$ such that $\framedtrans{(\Psi,\emptyset)}{\semb{P}}{\alpha}{P''}$ and $\semb{P'} \equiv P''$.
    \item If $\prioframedtrans{\Psi}{P}{\tau:p}{P'}$,
      then there exists $P''$ such that $\framedtrans{(\Psi,\emptyset)}{\semb{P}}{\tau}{P''}$ and $\semb{P'} \equiv P''$.
    \item If $\framedtrans{(\Psi,\emptyset)}{\semb{P}}{\alpha}{P'}$ and $\bn\alpha \freshin P$ and $\alpha \neq \tau$, then there exists $P''$ such that $\semb{P''} \equiv P'$ and $\prioframedtrans{\Psi}{P}{\alpha}{P''}$.
    \item If $\framedtrans{(\Psi,\emptyset)}{\semb{P}}{\tau}{P'}$, then there exists $p$ and $P''$ such that $\semb{P''} \equiv P'$ and $\prioframedtrans{\Psi}{P}{\tau:p}{P''}$.
  \end{enumerate}
\end{theorem}
Note that a simplification of the encoding with $\semb{\bang \alpha. P}  = \pass{(\infty)\alpha} \parop \bang \alpha. \semb{P} $ would render the above theorem false, since we would then lose the property that $\semb{\bang \alpha. P} \equiv \semb{\alpha. P \parop \bang\alpha. P}$, and transitions may unfold replications.

Gorla~\cite{DBLP:conf/concur/Gorla08} proposes a unified approach to encodability results, wherein a translation function is considered an encoding if it satisfies the five properties \emph{compositionality},
\emph{name invariance}, 
\emph{operational correspondence}, 
\emph{divergence reflection}, 
and
\emph{success sensitiveness}. 

Because our encoding satisfies strong operational correspondence,  the three last criteria  follow immediately. Name invariance is immediate since our encoding is equivariant, and compositionality holds with the caveat that we must consider replicated prefixes $!\alpha.P$ as an operator in itself, rather than considering the replication and the prefix as separate operators, and likewise for $\caseonly$-guarded prefixes.


Full abstraction means that two agents are equivalent iff their translations are equivalent.
The encoding presented in Section~\ref{sec:cuteencoding} is not fully abstract with respect to strong bisimilarity. This is because we require bisimilar agents to be statically equivalent, but the translation function introduces assertions such that the translation of bisimilar agents may not be statically equivalent. For a simple example, consider the agents $P = \alpha. \nil$ and $Q = \alpha. P$, where $\alpha$ is an output prefix. Clearly $P \parop P \priobisim Q$ holds, but for $\semb{P} = \pass{\alpha} \parop \alpha. (\nil \parop \pass{-\alpha})$ and $\semb{Q} = \pass{\alpha} \parop {\alpha. (\semb{P} \parop \pass{-\alpha})}$, we have $\frameof{\semb{P \parop P}} \vdash' (2)\alpha$ but $\frameof{\semb{Q}} \not\vdash' (2)\alpha$ and hence $\semb{P \parop P} \not\bisim \semb{Q}$.

At first glance, this difference between $\semb{P \parop P}$ and $\semb{Q}$ seems to be an unimportant technicality: the conditions $(2)\alpha$ and $(1)\alpha$ are not intended to be used as guards in $\caseonly$-statements. Their only use is in the evaluation of channel equivalences, but $\frameof{\semb{P \parop P}}$ and $\frameof{\semb{Q}}$ entail the same channel equivalences since the set of prefixes available coincides.
To motivate that they must be considered different, consider the distinguishing context $R = \pass{-\alpha} \parop \pass{\beta} \parop \gamma.0$, where $\beta$ is an input that can synchronise with $\alpha$, and $\gamma$ has lower priority than $\alpha$; we have that $R \parop \semb{Q}$ can take an action on $\gamma$, but $R \parop \semb{P \parop P}$ cannot. This highlights an interesting difference between $\mathcal{P}$ and $\mathcal{Q}$: in $\mathcal{P}$, a prefix describes both an interaction possibility and a constraint on other (lower-priority) interactions; in $\mathcal{Q}$, the interaction possibility and interaction constraint are two separate syntactical elements. This means that in $\mathcal{Q}$ we may write $\pass{\alpha}$, which is a process with no transitions that blocks lower-priority transitions as though it had an $\alpha$-transition; conversely $\alpha. P$ has a non-blocking $\alpha$-transition that may be blocked by higher-priority transitions.

Note that in the counterexample to full abstraction presented above, the context $R$ is not in the range of $\semb{\cdot}$. Thus our encoding may well satisfy weak full abstraction~\cite{Parrow_2008}, meaning that full abstraction holds if we restrict attention to contexts in the range of $\semb{\cdot}$. An investigation of this is deferred to future work.



A related question is whether a fully abstract encoding of $\mathcal{P}$ into some psi-calculus exists. The following theorem, inspired by recent work by Gorla and Nestmann~\cite{GorlaNestmann:Abstraction} and Parrow~\cite{Parrow:Abstraction}
, shows that because of the generality of the psi-calculi framework a trivial fully abstract ``encoding'' with strong bisimilarity as the target equivalence always exists, regardless of the source language and source equivalence under consideration.


Let $\mathbf{S}$ be a set ranged over by $s$, and $\sim$ be an equivalence on $\mathbf{S}$. Then
there is a psi-calculus $\mathcal S$ with no terms, with elements of $\mathbf{S}$ as assertions and conditions,  where entailment is $\sim$. Define the encoding $\semb{\_}_\mathbf{S} : \mathbf{S} \Rightarrow \processesof{\mathcal S}$ by $\semb{s}_\mathbf{S} \defn \pass{s}$.

\begin{theorem}
$s \sim s'$ iff $\semb{s}_\mathbf{S} \bisim \semb{s'}_\mathbf{S}$
\end{theorem}



This ``encoding'' simply embeds both the source language and source equivalence into a target language with no transition behaviour at all.
We conclude that 
a meaningful approach to full abstraction would have to impose additional criteria.
For an example, if we consider Gorla's criteria presented earlier, this ``encoding'' satisfies name invariance and divergence reflection, but fails to satisfy compositionality, operational correspondence and success sensitiveness.

\section{Conclusion}

In this paper, we have defined an extension of the psi-calculi framework with dynamic action priorities, and translated it to the original framework.
This illustrates the high expressiveness of the assertion mechanism in psi-calculi: usually, it is necessary to introduce negative premises or define a multi-layered transition system in order to obtain action priorities in a given calculus; for psi-calculi, what is already there suffices.

The extension with explicit priorities is interesting in its own right despite the encoding. Expressiveness is not usefulness. Modelling a system with priorities in terms of the translation would be more cumbersome than representing priorities directly.
Also, strong bisimulation in the extension is useful for proving equivalences that fail to hold in the encoding.

The most closely related development to psi-calculi with priorities is the \emph{attributed {\pic} with priorities}, written $\pi(\mathcal{L})$~\cite{articlereference201003038754456007}. It is designed as a generalisation of $\pi @$~\cite{DBLP:conf/esop/Versari07} and the stochastic {\pic}. Input and output prefixes take the form $e_1[e'_1]?\ve{x}$ and $e_2[e'_2]!\ve{y}$, where $e_1$ and $e_2$ are subjects, $\ve{x}$ and $\ve{y}$ are objects and $e'_1$ and $e'_2$ are interaction constraints, which may be instantiated to priorities or stochastic rates. $e$ ranges over expressions in an \emph{attribute language}, which is a kind of call-by-value $\lambda$-calculus equipped with a big-step reduction relation. The idea in the case of priorities is that if the expressions $e_1$ and $e_2$ reduce to the same channel name, and $\ve{e}$ reduces to some values $\ve{v}$, and the application $e'_1 e'_2$ reduces to the priority level $r$, then $e_1[e'_1]?\ve{x}.P \parop e_2[e'_2]!\ve{e}.Q$ reduces to $P\subst{\ve{x}}{\ve{v}} \parop Q$, unless another pair of prefixes can similarly communicate on a higher priority level. The focus is on developing type systems to prevent mismatches, on showing how the calculus can be applied to model phenomena in systems biology, and on the development and implementation of a stochastic simulation algorithm.

While $\pi(\mathcal{L})$ and our approach both generalise $\pi @$, the way the priorities are set up have several interesting differences that suggest incomparable expressive power in general. Priority levels in $\pi(\mathcal{L})$ are taken from an arbitrary partial order, whereas our priorities are natural numbers. Thus in $\pi(\mathcal{L})$ we may have systems where actions have mutually incomparable priority levels, unlike psi-calculi with priorities. The reason we use natural numbers is that the proof of Theorem~\ref{thm:priobisim} uses induction and successor arithmetic on the priority level; for future work we would like to investigate alternative proof strategies that would permit a generalised notion of priorities. In psi-calculi, priority levels are associated to communication channels, whereas in $\pi(\mathcal{L})$ they are associated with a particular pair of prefixes. The priority level of a particular pair in $\pi(\mathcal{L})$ is however static and cannot be influenced by the environment in any way, whereas in our approach priorities are dynamic and may change arbitrarily as the assertion environment evolves. While psi-calculi has no explicit notion of computation on data such as that given by the attribute language, the substitution function can be chosen so that it performs explicit computation on data, or implicit computation can be performed during the evaluation of entailments. For a detailed discussion of how to express computation on data in psi-calculi we refer to~\cite{Borgstr_m_2014}.

The translation assumes separate choice and prefix-guarded replication. An interesting question is if these assumptions can be relaxed. Allowing mixed choice is possible, but a different definition of guarding elements must be made, that records which prefixes occur in different branches of the same $\caseonly$-statement. With the current definition, $\semb{\caseonly\;{\ci{\top}{\overline{M}}\casesep \ci{\top}{\underline{M}}}}$ has the same guarding elements as $\semb{\overline{M} \parop \underline{M}}$, meaning that the former erroneously blocks other transitions as if a communication on $M$ could be derived. Allowing unguarded choice and replication would be more difficult, but we conjecture that it is possible at the expense of compositionality. The solution would involve extending the guarding elements to contain whole syntax trees, including binders. We then lose compositionality since if e.g. $\semb{\caseonly\;{\ci{\top}{P}\casesep\ci{\top}{Q}}}$ takes a transition from $Q$, the derivative must contain an assertion that retracts all interaction possibilities offered by $P$. Hence the translation of $Q$ depends on $P$, violating compositionality.

Another way to introduce priorities in process calculi is with a \emph{priority choice} operator $P +\rangle Q$, as is done for CCS in~\cite{Camilleri1991}. It is like the standard choice operator, with two exceptions. First, $P$ and $Q$ may for technical reasons not contain unguarded output prefixes. Second, transitions from $P$ take precedence over $Q$. More precisely, its semantics is defined so that it may always act as $P$, but may act as $Q$ only if no synchronisation on the prefixes of $P$ is possible in the current environment. This operator could be encoded in psi-calculi using techniques similar to those presented in this paper. The main idea is to augment the assertions with information about output prefixes as in Section~\ref{sec:cuteencoding}, and to translate priority choice as $\semb{P +\rangle Q} = \caseonly\;{\ci{\top}{\semb{P}} \casesep \ci{\varphi_P}{\semb{Q}}}$, where $\varphi_P$ is a condition that holds if no output prefixes matching the inputs of $P$ are enabled in the current environment. A more detailed investigation of this idea is deferred to future work.

We would also like to investigate if a result by Jensen~\cite{Torp94interpretingbroadcast}, that broadcast communication can be encoded in CCS with priority choice up-to weak bisimulation, can be adapted to our setting. If broadcast psi-calculi~\cite{borgstroem.huang.ea:broadcast-psi-sefm} can be encoded in psi-calculi with priorities, then by transitivity so can the original psi-calculi. This would contrast with the situation in the {\pic}, where broadcast communication cannot be encoded~\cite{ene.muntean:expressiveness-point}.

Since both the original psi-calculi and their extension with priorities have been formalised in Nominal Isabelle, we aim to formalise the correspondence results in this paper, in order to be more certain of their correctness. As a first step, it would be necessary to develop a formalisation of FIMMs in Isabelle, and integrate it with the nominal logic package.


\bibliographystyle{eptcs}
\bibliography{bibliography,pi,relatedBibliography}

\end{document}